\documentclass[11pt]{article}

\usepackage[preprint]{acl}

\usepackage{times}
\usepackage{url}
\usepackage{latexsym}
\usepackage[T1]{fontenc}
\usepackage[utf8]{inputenc}
\usepackage{microtype}
\usepackage{inconsolata}
\usepackage{graphicx}
\usepackage{fontawesome}
\usepackage{subfigure}
\usepackage{stfloats}
\usepackage{placeins}

\usepackage{amsmath,amssymb,amsthm}
\usepackage{booktabs}
\usepackage{array}
\usepackage{multirow}
\usepackage{enumitem}
\usepackage{xcolor}
\usepackage{algorithm}
\usepackage{algpseudocode}
\usepackage{tikz}
\usetikzlibrary{positioning,arrows.meta,fit,backgrounds,shapes.geometric}
\usepackage{CJKutf8}

\newcommand{\sj}{\textsc{SearchJev}}
\newcommand{\zh}[1]{\begin{CJK}{UTF8}{gbsn}#1\end{CJK}}

\newtheorem{proposition}{Proposition}

\title{SearchJev: A Calibrated System-1 Model for Search Agents}
\author{
  \textbf{Congfeng Cao}\textsuperscript{\rm 1} \quad
  \textbf{Lipeng Zuo}\textsuperscript{\rm 1} \quad
  \textbf{Konstantinos Papakostas}\textsuperscript{\rm 1} \quad
  \textbf{Qiwei Xu}\textsuperscript{\rm 1} \quad
  \textbf{Songwei Xu}\textsuperscript{\rm 1} \\
  \textbf{Qianru Meng}\textsuperscript{\rm 1} \quad
  \textbf{Lun Zhou}\textsuperscript{\rm 1} \quad
  \textbf{Zhaochun Ren}\textsuperscript{\rm 2} \quad
  \textbf{Yougang Lyu}\textsuperscript{\rm 1}\thanks{\ Corresponding authors.} \quad
  \textbf{Xiaohui Yan}\textsuperscript{\rm 1}\footnotemark[1] \\
  \textsuperscript{\rm 1}Huawei Technologies Co., Ltd. \qquad
  \textsuperscript{\rm 2}Leiden University \\[6pt]
  \normalsize
  \raisebox{-0.1em}{\faGithub}~\href{https://github.com/EvoScientist/SearchJev}{Code} \quad
  \raisebox{-0.2em}{\includegraphics[height=1em]{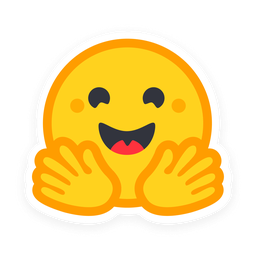}}~\href{https://huggingface.co/SearchJev/models}{Models}
}

\begin{document}
\maketitle


\begin{abstract}
Search agents repeatedly make short decisions about relevance, evidence
sufficiency, and search actions. Using generative language models for
these decisions introduces latency and unreliable confidence.
We present \sj{}, a calibrated System-1 model that separates
search decisions from System-2 reasoning and generation.
Given a search state and a decision schema, \sj{} scores legal
options without autoregressive output generation.
We propose Soft-Label Learning for Calibrated Decisions (SLCD)
to learn decision probabilities from uncertain supervision and improve the reliability of decision confidence.
In a dual-system search agent, \sj{} handles short decisions and delegates
uncertain judgments to System~2, which retains planning, query generation,
and answer composition.
We also introduce SearchDecision-Bench, a benchmark unifying six types of
search decisions for training and evaluation.
On SearchDecision-Bench, \sj{} improves decision quality over same-size
Qwen3.5 autoregressive models, achieves 5.2-5.3$\times$ faster decisions,
and reduces average expected calibration error by 41-74\%.
On BrowseComp-Plus, the dual-system agents achieve a 3.7-4.7$\times$
speedup in active search time while improving answer accuracy from 45\%
to up to 54\%. The benchmark and code are available\footnote{\url{https://github.com/EvoScientist/SearchJev}}.
\end{abstract}
\begin{figure}[!t]
  \centering
  \renewcommand{\subcapsize}{\fontsize{10}{12}\selectfont}
  \setlength{\subfigtopskip}{0pt}
  \setlength{\subfigcapskip}{0pt}
  \setlength{\subfigbottomskip}{0pt}
  \setlength{\abovecaptionskip}{4pt}
  \subfigure[Search decision performance.]{%
    \includegraphics[width=\columnwidth]{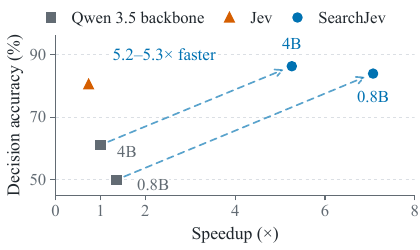}%
    \label{fig:pareto-decision}}
  \par\vspace{8pt}
  \subfigure[End-to-end agentic search.]{%
    \includegraphics[width=\columnwidth]{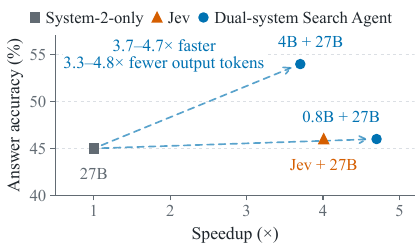}%
    \label{fig:pareto-search}}
  \caption{Search decision performance and end-to-end agentic search.
  (a) On SearchDecision-Bench, \mbox{\sj{}} achieves
  \mbox{5.2-5.3$\times$} faster decisions than same-size Qwen3.5.
  The x-axis uses Qwen3.5-4B as the 1$\times$ reference.
  (b) On BrowseComp-Plus, dual-system agents with
  \mbox{\sj{}-0.8B/4B} reach
  46.0\%/54.0\% answer accuracy, compared with 45.0\% for
  \mbox{System-2 only} and 46.0\% for Jev~1.13, a closed
System-1 model~\citep{typesafe2026jev}.
  Relative to System-2 only, they achieve 3.7-4.7$\times$ faster search
  with 3.3-4.8$\times$ fewer output tokens.}
  \label{fig:pareto}
\vspace*{-5mm}  
\end{figure}

\section{Introduction}
\label{sec:intro}

LLM-based search agents interleave reasoning with retrieval to answer
complex questions~\citep{yao2023react,jin2025searchr1,li2025searcho1}.
Within these trajectories, agents repeatedly make short decisions:
\emph{Is this passage relevant? Is the evidence sufficient to answer?
Which link should be explored next?}
These judgments determine which evidence is used and how the search
proceeds. Drawing on the distinction between System~1 and
System~2~\citep{kahneman2011thinking}, we view them as \emph{System-1}
decisions alongside the \emph{System-2} reasoning needed for planning
and answer generation. Handling both through the same generative process
adds sequential decoding overhead to frequent search decisions.

Two common approaches illustrate the difficulty of unifying search
decisions. Task-specific rerankers~\citep{nogueira2020monot5} and
query-complexity classifiers~\citep{jeong2024adaptiverag} support individual
functions, but require separate models and supervision for different
decisions. Prompting a general LLM accommodates diverse tasks, but uses
autoregressive generation to produce short judgments. Its confidence,
whether verbalized or derived from token probabilities, is often
miscalibrated~\citep{kadavath2022language,tian2023just,xiong2024can}.
Reliable confidence is needed to determine whether a decision can be
accepted or requires reasoning. The challenge is therefore to
handle heterogeneous search decisions with a unified model that is both
efficient and well calibrated.


To address this challenge, we propose \sj{}, a fast and calibrated
System-1 model for search agents. Inspired by Jev's probabilistic
outputs for automation~\citep{typesafe2026jev}, we formulate search judgments as schema-conditioned probabilistic
decisions, separating them from System-2 reasoning and generation
(see Appendix~\ref{app:related} for related work).
This allows one
model to handle multiple decision types and provide confidence
estimates for deciding when to invoke System~2.

\sj{} has three components.
(i) The \emph{System-1 Search Decision Model} directly scores legal
options from the LM's first-output-token logits, conditioned on the
search state and a textual decision schema, avoiding autoregressive
output generation.
(ii) \emph{Soft-Label Learning for Calibrated Decisions} (SLCD) learns
decision distributions from uncertain supervision using soft targets
and proper scoring losses, followed by temperature calibration to
improve confidence and generalization to held-out data
~\citep{gneiting2007strictly,guo2017calibration}.
(iii) The \emph{Dual-System Search Agent} accepts confident \sj{}
decisions and delegates uncertain judgments to System~2, which retains
planning, query generation, and answer composition.

To support unified training and evaluation, we introduce
SearchDecision-Bench. Existing supervision for search decisions comes
from heterogeneous resources with different input and label formats.
We convert these resources into a common state-schema format covering
six decision types: routing, rewriting, relevance, sufficiency,
navigation, and verification. The benchmark evaluates decision quality,
calibration, and latency, and includes held-out datasets and splits for
assessing generalization.

We evaluate \sj{} at both the decision and agent levels.
On SearchDecision-Bench, \sj{} improves decision quality and reduces
average calibration error relative to same-size Qwen3.5 AR (JSON)
models, while achieving 5.2-5.3$\times$ faster decisions.
On BrowseComp-Plus~\citep{chen2026browsecompplus}, the dual-system agents
achieve a 3.7-4.7$\times$ speedup in median active time while improving
answer accuracy from 45\% to up to 54\%, compared with the System-2 only
agent. Figure~\ref{fig:pareto} summarizes these gains in search decision
performance and end-to-end agentic search.

Our contributions are:
\begin{itemize}[leftmargin=*,nosep]
  \item We propose \sj{}, a schema-conditioned search decision model that
        directly predicts legal options and learns decision probabilities
        through SLCD.
  \item We build a dual-system search agent that separates short decisions
        from reasoning and generation, and delegates uncertain judgments
        from System~1 to System~2.
  \item We introduce SearchDecision-Bench, unifying six search decision types.
  \item We demonstrate improved decision quality, average calibration,
        and efficiency over same-size autoregressive models, and better
        end-to-end answer accuracy with lower search time than a
        System-2 only agent.
\end{itemize}
\section{SearchDecision-Bench}
\label{sec:bench}

We introduce SearchDecision-Bench to train and evaluate short decisions
made during agentic search. It unifies heterogeneous search resources as
state-schema pairs with field-level targets, covering six decision types
and three output types.

\subsection{Task Definition and Taxonomy}
\label{sec:problem}

\paragraph{Task definition.}
Given a search state and a decision schema, the task is to predict a legal
option for each requested decision and estimate its confidence.
At step $t$, the state $s_t=(q,H_t,C_t)$ contains the user request $q$,
the trajectory $H_t$ of queries, opened documents, and intermediate notes,
and candidates $C_t=\{c_1,\dots,c_n\}$ such as documents, links, or sources.
A schema $\sigma=\{f_1,\dots,f_k\}$ specifies the decisions as named
fields, each with a natural-language question and a set of legal options
$\mathcal{V}(f)$. A sample consists of a serialized state, a schema, and
a target for each field; several fields can share the same state.
We write $s=s_t$ when the step index is not needed.

\begin{table*}[!t]
  \centering
  \footnotesize
  \setlength{\tabcolsep}{3pt}
  \setlength{\arrayrulewidth}{0.3pt}
  \renewcommand{\arraystretch}{0.9}
  \newcommand{\slot}[1]{\textit{\{#1\}}}
  \newcommand{\taskblock}[1]{%
    \begin{tabular}[c]{@{}>{\raggedright\arraybackslash}m{2.8cm}
      >{\raggedright\arraybackslash}m{2.0cm}
      >{\raggedright\arraybackslash}m{\dimexpr\textwidth-6.7cm-6\tabcolsep\relax}@{}}
      #1
    \end{tabular}}
  \newcommand{\tasksep}{\\[0.5pt]\noalign{%
    \hbox to \dimexpr\textwidth-1.9cm-2\tabcolsep\relax{%
      \xleaders\hbox{\rule{1.5pt}{\arrayrulewidth}\hskip2pt}\hfill}%
    \vskip0.5pt}}
  \begin{tabular}{@{}>{\raggedright\arraybackslash}m{1.9cm}l@{}}
    \toprule
    \textbf{Decision Type} & \taskblock{%
      \textbf{Decision Task} & \textbf{Output Type} & \textbf{Task Example}} \\
    \midrule
    Routing & \taskblock{%
      Intent classification & Choice
      & Query: \slot{query}. Which intent does it express?\newline
        Options: \slot{intent labels}.
      \tasksep
      Retrieval need & Score
      & Query: \slot{query}. How much information must be retrieved?\newline
        Levels: none / one fact / several facts / multi-source research.
      \tasksep
      Query clarity & Choice
      & Query: \slot{query}. What information most needs to be added?\newline
        Options: none / time / location / object / aspect / \slot{other}.} \\
    \midrule
    Rewriting & \taskblock{%
      Query equivalence & Noul
      & Queries: \slot{query 1} and \slot{query 2}.\newline
        Do they express the same intent? Options: True / False.
      \tasksep
      Rewrite equivalence & Noul
      & Dialogue: \slot{history}. Utterance: \slot{utterance}. Rewrite: \slot{rewrite}.\newline
        Does the rewrite keep the utterance's intent? Options: True / False.} \\
    \midrule
    Relevance & \taskblock{%
      Relevance judgment & Noul / Score
      & Query: \slot{query}. Document: \slot{document}.\newline
        Is it relevant? Options: True / False, or \slot{relevance grades}.
      \tasksep
      Document usefulness & Noul
      & Sub-query: \slot{current sub-query}. Document: \slot{document}.\newline
        Does the document answer the sub-query? Options: True / False.
      \tasksep
      Document quality & Score
      & Query: \slot{query}. Document: \slot{site, date, content}.\newline
        How authoritative is the source? Levels: untrusted / user-generated /
        professional / authoritative.} \\
    \midrule
    Sufficiency & \taskblock{%
      Evidence sufficiency & Noul
      & Question: \slot{original question}. Evidence: \slot{retrieved evidence}.\newline
        Does the evidence cover all facts needed to answer?\newline
        Options: True / False.} \\
    \midrule
    Navigation & \taskblock{%
      Next-action selection & Choice
      & Search state: \slot{state}. What should the agent do next?\newline
        Options: search / read page / answer.
      \tasksep
      Next sub-query selection & Choice
      & Question: \slot{original question}. Solved steps: \slot{history}.\newline
        Which sub-query comes next? Options: \slot{four candidate queries}.
      \tasksep
      Query-fit assessment & Noul
      & Next unsolved step: \slot{step}. Candidate query: \slot{query}.\newline
        Does the query target this step? Options: True / False.} \\
    \midrule
    Verification & \taskblock{%
      Claim verification & Choice
      & Claim: \slot{claim}. Evidence: \slot{evidence}.\newline
        Is the claim supported? Options: supports / refutes / insufficient.
      \tasksep
      Evidence detection & Noul
      & Claim: \slot{claim}. Sentence: \slot{sentence}.\newline
        Is the sentence evidence for the claim? Options: True / False.} \\
    \bottomrule
  \end{tabular}
  \vspace{-1mm}
  \caption{Representative decision tasks for agentic search in SearchDecision-Bench.
  The illustrative examples span six decision types and three output types:
  Choice (categorical selection), Score (ordinal scoring), and Noul (binary judgment).
  Braces denote placeholders.}
  \label{tab:taxonomy}
  \vspace{-2mm}
\end{table*}

\begin{table}[!t]
  \centering
  \small
  \setlength{\tabcolsep}{3pt}
  \begin{tabular*}{\columnwidth}{@{}l@{\extracolsep{\fill}}rrr@{}}
    \toprule
    \textbf{Decision Type} & \textbf{Training} & \textbf{ID Test} & \textbf{OOD Test} \\
    \midrule
    Routing      &  77,627 & 26,427 & 10,000 \\
    Rewriting    &  74,318 & 32,746 & 10,000 \\
    Relevance    & 153,990 & 17,581 &  5,976 \\
    Sufficiency  &  14,652 &  1,036 &      0 \\
    Navigation   &  56,981 & 10,213 &      0 \\
    Verification &  47,830 & 10,334 & 20,631 \\
    \bottomrule
  \end{tabular*}
  \vspace{-1mm}
  \caption{Statistics of SearchDecision-Bench.
  Counts denote state-schema pairs.}
  \label{tab:bench}
  \vspace{-2mm}
\end{table}

\paragraph{Decision types.}
We organize search decisions into six types: \emph{routing} identifies a
query's intent or retrieval needs; \emph{rewriting} checks a candidate
query rewrite; \emph{relevance} judges retrieved documents;
\emph{sufficiency} determines whether the available evidence is enough;
\emph{navigation} selects or assesses the next search step; and
\emph{verification} checks whether evidence supports a claim.
Table~\ref{tab:taxonomy} presents representative tasks and examples for
each type.

\paragraph{Output types.}
SearchDecision-Bench uses \emph{Choice} for categorical selection,
\emph{Score} for ordinal scoring, and \emph{Noul} for binary judgments following Jev~\citep{typesafe2026jev}.
Legal options are specified in text; a decision type can use multiple
output types.

\subsection{Data Construction and Splits}
\label{sec:bench-data}

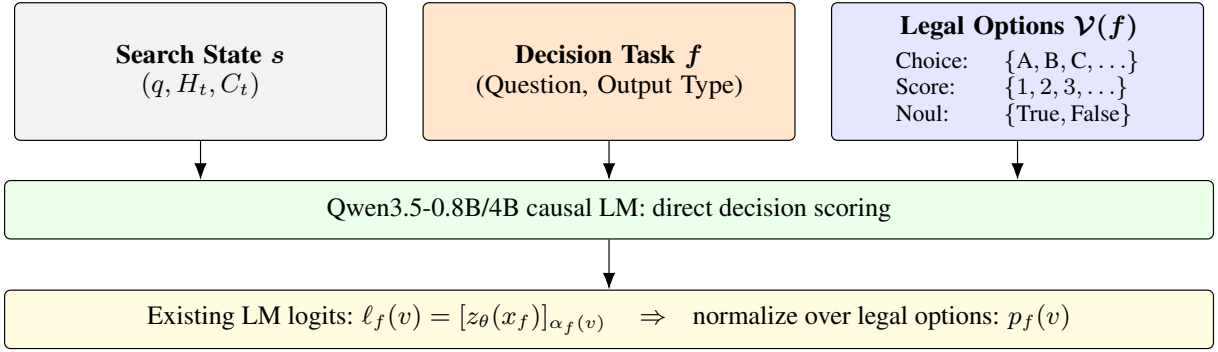
\begin{figure*}[!t]
  \centering
  \resizebox{\textwidth}{!}{%
  \begin{tikzpicture}[
      font=\small,
      box/.style={draw, rounded corners=2pt, align=center, minimum height=0.7cm, inner sep=3pt},
      tok/.style={box, fill=blue!10},
      arr/.style={-{Latex[length=2mm]}}]
    \node[box, fill=gray!10, text width=4.25cm, minimum height=1.65cm] (state) at (-4.9,0)
      {\textbf{Search State $\boldsymbol{s}$}\\$(q, H_t, C_t)$};
    \node[tok, fill=orange!20, text width=4.25cm, minimum height=1.65cm] (field) at (0,0)
      {\textbf{Decision Task $\boldsymbol{f}$}\\(Question, Output Type)};
    \node[tok, text width=4.25cm, minimum height=1.65cm] (options) at (4.9,0)
      {\textbf{Legal Options $\boldsymbol{\mathcal{V}(f)}$}\\[2pt]
       {\fontsize{8}{9}\selectfont
       \begin{tabular}{@{}ll@{}}
         Choice: & $\{\text{A},\text{B},\text{C},\ldots\}$\\
         Score: & $\{1,2,3,\ldots\}$\\
         Noul: & $\{\text{True},\text{False}\}$
       \end{tabular}}};
    \path (state.west) -- coordinate[midway] (mid) (options.east);
    \node[box, fill=green!8, minimum width=14.5cm, below=1.3cm of mid] (lm)
      {Qwen3.5-0.8B/4B causal LM: direct decision scoring};
    \node[box, fill=yellow!15, below=0.6cm of lm, minimum width=14.5cm] (readout)
      {Existing LM logits: $\ell_f(v)=[z_\theta(x_f)]_{\alpha_f(v)}$
       \quad$\Rightarrow$\quad normalize over legal options: $p_f(v)$};
    \draw[arr] (state.south) -- (state.south |- lm.north);
    \draw[arr] (field.south) -- (field.south |- lm.north);
    \draw[arr] (options.south) -- (options.south |- lm.north);
    \draw[arr] (lm) -- (readout);
  \end{tikzpicture}%
  }
  \vspace{-1mm}
  \caption{\sj{} decision model. A search state, the question and output type
  of task $f$, and its legal options form the decision input $x_f$.
  Choice uses letter labels; Score and Noul show example values. The model normalizes
  option-label logits into a decision distribution without autoregressive
  output generation.}
  \label{fig:arch}
  \vspace{-2mm}
\end{figure*}

\paragraph{Data sources.}
We draw on Chinese and English resources that provide supervision for
complementary search decisions. Query-understanding and rewriting datasets,
such as BANKING77~\citep{casanueva2020efficient},
LCQMC~\citep{liu2018lcqmc}, and QReCC~\citep{anantha2021qrecc}, support
intent classification and query-equivalence decisions. Retrieval datasets, including
T2Ranking~\citep{xie2023t2ranking} and
DuReader-Retrieval~\citep{qiu2022dureader}, provide passage-level judgments.
Multi-hop QA and agent demonstrations support sufficiency and navigation
decisions, while claim-evidence resources such as
VitaminC~\citep{schuster2021vitaminc} provide verification labels.
Training also includes natural language inference, sentiment analysis,
topic classification, and reading comprehension
data~\citep{xu2020clue,tan2008empirical}.

\paragraph{Sample construction and annotation.}
We convert each source instance into a search state and instantiate
decision fields with textual questions and legal options. States range from
query-passage, query-query, and claim-evidence pairs to multi-hop states
built from MuSiQue~\citep{trivedi2022musique} decompositions.
Supervision comes from original annotations, derived labels, and LLM
teachers or votes. For example, MuSiQue decompositions and hard negatives
provide evidence-sufficiency and next-step judgments, filtered by an LLM;
agent demonstrations provide next-action labels.
We encode labels as soft targets, assigning confidence according to the
supervision source, and paraphrase field descriptions in Chinese or English
to express the same decision through different instructions.

\paragraph{Quality control.}
We deduplicate identical inputs, remove samples with conflicting labels,
and mask fields that cannot be judged from the available information.
We isolate splits using identity keys and query text, and discard synthetic
variants whose source appears in an evaluation split.
Training and validation inputs whose states also occur in a test set are
removed.


\paragraph{Splits and statistics.}
In-distribution (ID) tests use held-out samples from training datasets.
Out-of-distribution (OOD) evaluation uses held-out datasets
and splits: FRAMES~\citep{krishna2025frames}, RAGTruth~\citep{niu2024ragtruth},
CANDY~\citep{guo2025candy}, Search Arena~\citep{miroyan2026searcharena},
AgentRewardBench~\citep{lu2025agentrewardbench}, and the QReCC test split.
ID and OOD test samples are excluded from training.
Table~\ref{tab:bench} counts state-schema pairs across six search-decision
types: all six have ID tests, and four have OOD tests.
General-language training tasks are outside these counts.
BrowseComp-Plus~\citep{chen2026browsecompplus} is used only for
end-to-end agentic search evaluation.

\section{\sj{}}
\label{sec:method}

\sj{} separates short search decisions from System~2 reasoning and generation.
It directly predicts calibrated distributions over legal options without
autoregressive decoding, delegating uncertain decisions to System~2,
which handles planning, query generation, and answer synthesis.

\subsection{System-1 Search Decision Model}
\label{sec:model}

\sj{} predicts search decisions from the output logits of a pretrained
causal LM~\citep{qwen2026qwen35}, as illustrated in
Figure~\ref{fig:arch}.

\begin{figure*}[t]
    \vspace{-1mm}
    \centering
    \includegraphics[width=\textwidth]{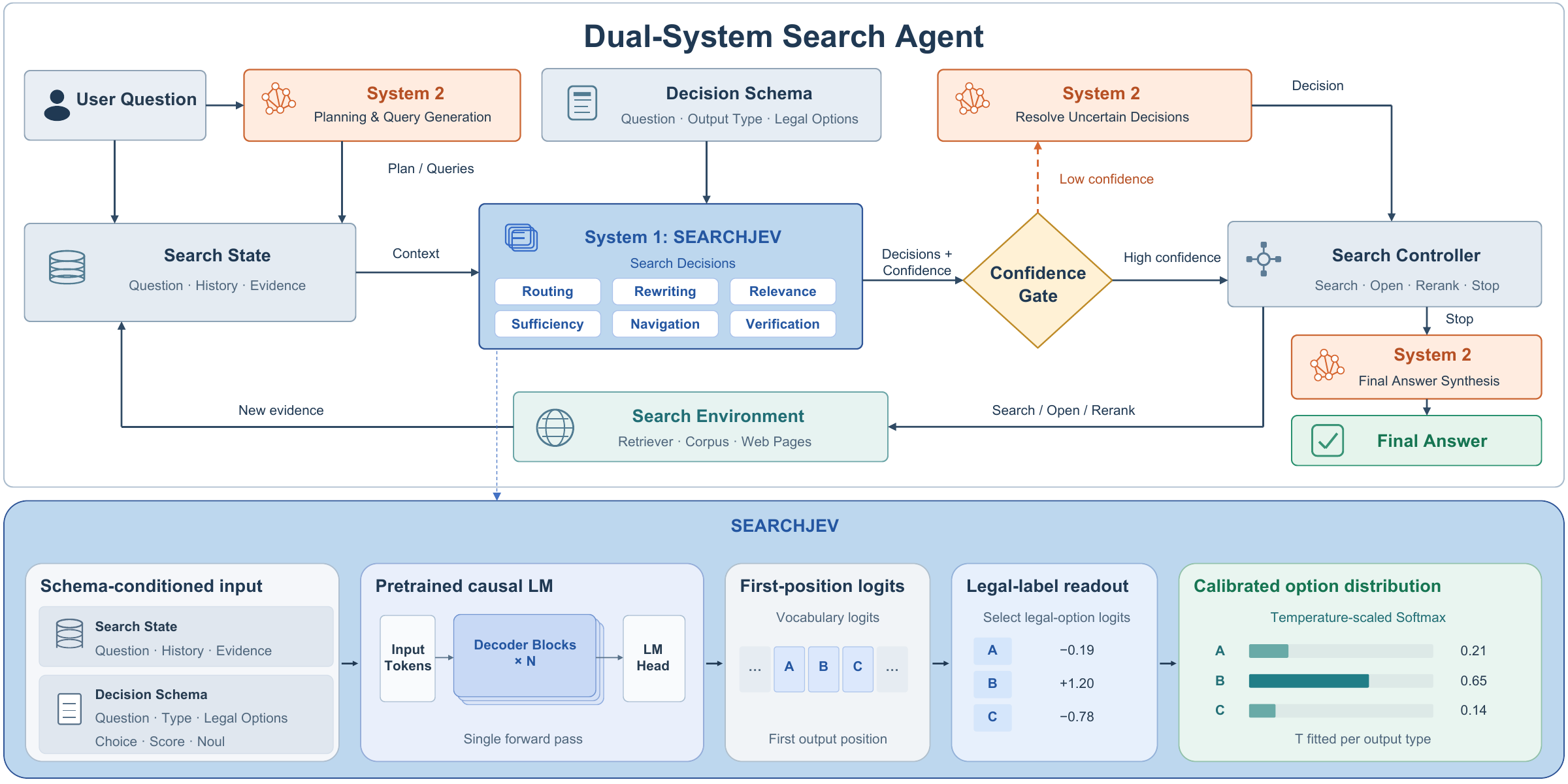}
    \caption{Architecture of the Dual-System Search Agent.
  System~2 handles planning, query generation, and final answer synthesis,
  while \sj{} (System~1) makes search decisions specified by the schema at
  each step. Confident decisions are accepted; uncertain ones are delegated
  to System~2. The search controller uses the resolved decisions to select
  actions and update the search state with retrieved evidence.}
    \label{fig:overview}
    \vspace{-1mm}
\end{figure*}

\paragraph{Schema-conditioned input.}
For each task $f\in\sigma$, the input $x_f$ contains the search state $s$,
the task question, output type, and legal options $\mathcal{V}(f)$.
Each option is assigned a distinct single-token label (e.g., A or B),
although its description may span multiple tokens.
The input ends with a label-selection instruction and the assistant
generation prefix.

\paragraph{Token-logit decision readout.}
Let $z_\theta(x_f)$ denote the LM logits at the first output position,
and $\alpha_f(v)$ the token index assigned to option $v$.
The option score and uncalibrated probability are
\begin{equation}
  \begin{aligned}
  \ell_f(v) &= [z_\theta(x_f)]_{\alpha_f(v)},\\
  p_f(v) &= \frac{\exp\ell_f(v)}
  {\sum_{u\in\mathcal{V}(f)}\exp\ell_f(u)},
  \end{aligned}
  \label{eq:logit}
\end{equation}
where $v\in\mathcal{V}(f)$ and $p_f(v)=p_\theta(v\mid x_f)$.
Normalizing over legal labels yields distributions for Choice, Score,
and Noul decisions using the existing LM head without task-specific
parameters. This supports batched inference without autoregressive
decoding or JSON parsing.

\subsection{Soft-Label Learning for Calibrated Decisions}
\label{sec:training}

We propose \emph{Soft-Label Learning for Calibrated Decisions}
(SLCD) to learn decision probabilities from heterogeneous
supervision and calibrate confidence by output type.

\paragraph{Soft-label supervision.}
Each field uses a target distribution $t_f$ over $\mathcal{V}(f)$,
defaulting to one-hot labels when soft targets are unavailable.
Retaining label uncertainty improves held-out accuracy
over one-hot labels (Section~\ref{sec:generalization-results}).
We paraphrase field descriptions in Chinese or English and permute
unordered options to reduce position bias
~\citep{zheng2024large,pezeshkpour2024large}.

\paragraph{Decision probability learning.}
We combine cross-entropy and Brier losses:
\begin{equation}
  \begin{split}
  \mathcal{L}_f=w_f\Big[&-\sum_{v\in\mathcal{V}(f)} t_f(v)\log p_f(v)\\
  &+\lambda_B\sum_{v\in\mathcal{V}(f)}\big(p_f(v)-t_f(v)\big)^2\Big],
  \end{split}
  \label{eq:sft}
\end{equation}
where $\lambda_B\ge0$ balances the losses and $w_f>0$ downweights
LLM-assigned labels, demonstrations, and constructed rewrites.
We average the loss over fields in each minibatch.
Both losses are strictly proper scoring rules, whose combination
is minimized at $t_f$
~\citep{brier1950verification,gneiting2007strictly}.

\paragraph{Output-type temperature calibration.}
We fit a positive temperature for each output type by minimizing
validation negative log-likelihood~\citep{guo2017calibration}:
\begin{equation}
  \tilde p_f(v)=\frac{\exp(\ell_f(v)/T_{\kappa(f)})}
  {\sum_{u\in\mathcal{V}(f)}\exp(\ell_f(u)/T_{\kappa(f)})},
  \label{eq:calibration}
\end{equation}
where $\kappa(f)\in\{\text{Choice},\text{Score},\text{Noul}\}$.
Types with insufficient validation data share a pooled temperature.
Calibration preserves option rankings, and $\tilde p_f$
provides confidence estimates for the search agent.

\subsection{Dual-System Search Agent}
\label{sec:collab}
\label{sec:agent}

Figure~\ref{fig:overview} illustrates the dual-system search loop.
\sj{} (System~1) handles short, schema-defined search decisions,
while System~2 performs planning, query generation, and answer
synthesis, and resolves uncertain judgments.
A shared search controller translates the decisions into actions.

\paragraph{Schema-conditioned decisions.}
Given search state $s_t$ and schema $\sigma$, \sj{} processes
$\{x_f\}_{f\in\sigma}$ and returns calibrated distributions
$\{\tilde p_f\}_{f\in\sigma}$ over legal options.
Adapting the schema and controller rules supports different
search tasks while preserving the same division of labor.

\paragraph{Confidence-gated delegation.}
For each field $f$, the gate accepts the System-1 decision if
$\max_{v\in\mathcal{V}(f)}\tilde p_f(v)\ge\delta$;
otherwise, System~2 selects from the same legal options.
The threshold $\delta\in[0,1]$ controls delegation
~\citep{geifman2017selective}, with both paths feeding
the search controller.

\paragraph{Search execution and state update.}
The controller maps decisions to search, open, rerank, or stop
actions. Retrieved evidence updates $s_{t+1}$ for subsequent
decisions. Upon stopping, System~2 synthesizes the final answer.

\section{Experiments}
\label{sec:experiments}
\label{sec:eval-setup}

We investigate three research questions:

\noindent\textbf{RQ1}: How does \sj{} perform in terms of decision effectiveness and efficiency?

\noindent\textbf{RQ2}: How does \sj{} affect the answer quality and efficiency of end-to-end agentic search?

\noindent\textbf{RQ3}: How well does \sj{} generalize to held-out datasets across different decision families?


\subsection{Baselines}
\label{sec:setup}
\label{sec:baselines}
To evaluate the effectiveness of \sj{} at the decision and agent levels,
we compare it with the following baselines:
\begin{itemize}[leftmargin=*,nosep]
  \item \textbf{Qwen3.5 AR (JSON)} uses the same Qwen3.5-0.8B/4B
  backbones before task-specific training, with constrained JSON decoding
  on the same prompts. Confidence scores are derived from option
  log-probabilities.
  \item \textbf{\sj{} (one-hot)} trains the same Qwen3.5-0.8B/4B
  backbones with the same data, prompts, objective (Eq.~\ref{eq:sft}),
  and output-type temperature calibration as \sj{}, but replaces every
  soft target with the one-hot label of its reference option.
  \item \textbf{Jev 1.13}~\citep{typesafe2026jev} is a closed
  System-1 model, which we query through its API on the same test items.
  We also use Jev in place of \sj{} in the dual-system agent for
  end-to-end comparison.
  \item \textbf{System-2 only} uses Qwen3.5-27B for short decisions
  and search reasoning. We compare it with dual-system agents that
  delegate the short decisions to \sj{}-0.8B or \sj{}-4B.
\end{itemize}
\subsection{Evaluation Metrics}
\label{sec:metrics}
To evaluate decision effectiveness on SearchDecision-Bench,
we report accuracy for the five classification-based
families and NDCG@10~\citep{jarvelin2002cumulated} for relevance.
We measure calibration using expected calibration error
(ECE;~\citealp{naeini2015obtaining}) and efficiency using median latency
per decision.

For end-to-end agentic search, we follow the answer-evaluation protocol of
BrowseComp-Plus~\citep{chen2026browsecompplus}: an LLM judge compares the
answer with the reference answer, and accuracy is the percentage of
questions judged correct. We also report System-2 calls, output and
reasoning tokens, and median active time per question, including System-2
calls, decision-model inference, and retrieval.

\subsection{Implementation Details}
\label{sec:implementation}
We initialize \sj{}-0.8B and \sj{}-4B from Qwen3.5-0.8B/4B~\citep{qwen2026qwen35}
and train them with LoRA~\citep{hu2022lora} using SLCD.
We set $\lambda_B=0.5$ and assign $w_f=0.5$ to LLM-assigned labels,
demonstrations, and constructed rewrites, and $w_f=1$ otherwise.
Output-type temperatures are fitted on 5,000 validation decisions.
Decision latency is measured at batch size 1 using
the same inference code for the open models. See Appendix~\ref{app:Experimental Setting} for detail.

For end-to-end evaluation, we randomly sample 100 BrowseComp-Plus
questions and run each agent once per question, using Qwen3.5-27B as
System~2 and BM25~\citep{robertson2009probabilistic} as the retriever.
We consider three decision types in BrowseComp-Plus:
(i) \emph{relevance} judges retrieved documents against the current query
to select useful evidence;
(ii) \emph{sufficiency} determines whether the evidence is enough to answer
or further search is needed; and
(iii) \emph{verification} assesses whether the retained evidence supports
the final answer, using the calibrated support probability as an
answer-support score.

\section{Experimental Results and Analysis}
\label{sec:exp}
\label{sec:results}

\begin{table*}[!tb]
  \centering
  \small
  \setlength{\tabcolsep}{3pt}
  \resizebox{\textwidth}{!}{%
  \begin{tabular}{@{}l cc cc cc cc cc cc c@{}}
    \toprule
    & \multicolumn{2}{c}{Relevance} & \multicolumn{2}{c}{Sufficiency}
    & \multicolumn{2}{c}{Routing} & \multicolumn{2}{c}{Navigation}
    & \multicolumn{2}{c}{Rewriting} & \multicolumn{2}{c}{Verification} & Latency \\
    \cmidrule(lr){2-3}\cmidrule(lr){4-5}\cmidrule(lr){6-7}\cmidrule(lr){8-9}\cmidrule(lr){10-11}\cmidrule(lr){12-13}
    Model & NDCG$\uparrow$ & ECE$\downarrow$ & Acc$\uparrow$ & ECE$\downarrow$
          & Acc$\uparrow$ & ECE$\downarrow$ & Acc$\uparrow$ & ECE$\downarrow$
          & Acc$\uparrow$ & ECE$\downarrow$ & Acc$\uparrow$ & ECE$\downarrow$ & ms$\downarrow$ \\
    \midrule
    Jev 1.13 (closed)            & 98.5 & 7.6 & 79.8 & 5.6 & 73.5 & 5.5 & \textbf{72.1} & \textbf{1.5} & 79.5 & 2.8 & 79.8 & \textbf{1.6} & 265.4 \\
    \midrule
    Qwen3.5-0.8B AR (JSON)       & 85.7 & 15.7 & 49.9 & 24.2 & 24.7 & 24.0 & 42.0 & 19.7 & 40.6 & 32.1 & 56.8 & 9.5 & 144.2 \\
    Qwen3.5-4B AR (JSON)         & 96.7 & 4.2 & 57.7 & 15.0 & 53.3 & 6.7 & 37.8 & 26.3 & 66.1 & 5.5 & 55.1 & 4.7 & 195.2 \\
    \midrule
    \sj{}-0.8B (one-hot) & \underline{98.7} & \underline{3.9} & 89.9 & \underline{4.2} & 83.2 & \underline{4.3} & 56.6 & 5.7 & 84.0 & 5.2 & 91.9 & 4.5 & \textbf{27.6} \\
    \sj{}-4B (one-hot)   & \textbf{99.0} & \textbf{3.2} & \underline{95.5} & \textbf{3.4} & \underline{86.5} & \textbf{3.5} & 55.3 & 10.8 & \textbf{86.0} & 5.0 & \textbf{93.6} & \underline{4.2} & \underline{37.1} \\
    \midrule
    \textbf{\sj{}-0.8B}          & 98.6 & 5.8 & 89.4 & 5.1 & 83.2 & 5.0 & 56.8 & \underline{4.6} & 84.0 & \underline{2.3} & 91.7 & 9.3 & \textbf{27.6} \\
    \textbf{\sj{}-4B}            & \textbf{99.0} & 6.5 & \textbf{96.0} & 5.6 & \textbf{86.6} & 5.7 & \underline{57.2} & 8.4 & \underline{85.9} & \textbf{2.0} & \underline{93.2} & 8.8 & \underline{37.1} \\
    \bottomrule
  \end{tabular}}
  \vspace{-1mm}
  \caption{Main results on the in-distribution test sets of
  SearchDecision-Bench, covering decision quality, calibration, and latency.
  \sj{} (one-hot) is trained identically to \sj{} but with one-hot
  labels in place of soft targets (Section~\ref{sec:baselines}); its latency equals that of \sj{},
  which has the same architecture and inference code.
  NDCG@10, accuracy, and ECE are scaled by 100; latency is the median
  per-decision time in milliseconds at batch size 1.
  \textbf{Bold} highlights the best performance,
  \underline{underlined} indicates the second-best.}
  \label{tab:main}
  \vspace{-2mm}
\end{table*}

We evaluate \sj{} on decision effectiveness, end-to-end agentic
search, and generalization to held-out datasets.

\subsection*{RQ1: Decision Effectiveness and Efficiency}
\label{sec:decision-results}
\begin{table*}[!t]
  \centering
  \small
  \setlength{\tabcolsep}{2pt}
  \begin{tabular*}{\textwidth}{@{}l@{\extracolsep{\fill}}ccccc@{}}
    \toprule
    Agent & Acc $\uparrow$ & $\Delta$Acc & System-2 Calls $\downarrow$
          & Output Tokens (k) $\downarrow$ & Time (s) $\downarrow$ \\
    \midrule
    System-2 only (Qwen3.5-27B)  & 45.0 & -- & 90.3 & 73.7 & 4080 \\
    \midrule
    Dual-system Search Agent (Jev~1.13) & \underline{46.0} & -- & \underline{76.0} & \underline{18.3} & \underline{1018}  \\
    \midrule
    \textbf{Dual-system Search Agent (\sj{}-0.8B)} & \underline{46.0} & -- & \textbf{62.1} & \textbf{15.5} & \textbf{868} \\
    \multicolumn{6}{@{}l@{}}{\hbox to \textwidth{\xleaders\hbox to 4pt{\hss\rule[1.25ex]{2pt}{0.4pt}\hss}\hfill}}\\[-1.7ex]
    \quad Relevance by Qwen3.5-0.8B AR (JSON)    & 41.0 & $-$5.0 & 52.5 & 15.7 & 718 \\
    \quad Sufficiency by Qwen3.5-0.8B AR (JSON)  & 16.0 & $-$30.0 & 6.6  & 2.5  & 72 \\
    \quad Verification by Qwen3.5-0.8B AR (JSON) & 44.0 & $-$2.0 & 69.9 & 17.2 & 711 \\
    \midrule
    \textbf{Dual-system Search Agent (\sj{}-4B)}   & \textbf{54.0} & -- & 84.6 & 22.1 & 1102 \\
    \multicolumn{6}{@{}l@{}}{\hbox to \textwidth{\xleaders\hbox to 4pt{\hss\rule[1.25ex]{2pt}{0.4pt}\hss}\hfill}}\\[-1.7ex]
    \quad Relevance by Qwen3.5-4B AR (JSON)      & 53.0 & $-$1.0 & 88.8 & 21.6 & 1042 \\
    \quad Sufficiency by Qwen3.5-4B AR (JSON)    & 44.0 & $-$10.0 & 61.0 & 15.3 & 808 \\
    \quad Verification by Qwen3.5-4B AR (JSON)   & 45.0 & $-$9.0 & 86.6 & 19.6 & 913 \\
    \bottomrule
  \end{tabular*}
  \vspace{-1mm}
\caption{End-to-end agentic search results on BrowseComp-Plus
with System~2 fixed at Qwen3.5-27B.
Indented rows replace one \sj{} decision type with
same-size Qwen3.5 AR (JSON).
$\Delta$Acc denotes the accuracy change (percentage points)
relative to the corresponding \sj{} agent.
Relevance and sufficiency affect search control, while
verification is post-hoc; its accuracy differences reflect
run-to-run variation.
System-2 calls and output tokens (k: thousands) are per question;
time is median active time including inference and retrieval.
Bold and underline indicate the best and second-best
non-ablation results.}
  \label{tab:e2e}
  \vspace{-2mm}
\end{table*}

Table~\ref{tab:main} compares decision quality, calibration, and
latency across six in-distribution decision families.
We make four observations:
\begin{itemize}[leftmargin=*,topsep=2pt,itemsep=2pt,parsep=0pt]
  \item \textbf{\sj{} improves decision quality across all six families.}
  Both model sizes outperform same-size Qwen3.5 AR (JSON),
  with \sj{}-4B or its one-hot variant leading five families.
  Compared with Jev~1.13, both achieve higher accuracy on
  sufficiency, routing, rewriting, and verification,
  comparable relevance, but lower navigation accuracy.

  \item \textbf{\sj{} improves average decision calibration.}
  Mean ECE decreases by 74.4\% and 40.7\% relative to
  same-size Qwen3.5 AR (JSON) at 0.8B and 4B, respectively.
  Compared with Jev~1.13, \sj{}-0.8B achieves lower ECE
  on four families, while \sj{}-4B improves relevance
  and rewriting and matches sufficiency.
  Jev retains the lowest mean ECE, particularly on
  navigation and verification.

  \item \textbf{Soft targets match one-hot accuracy with better-behaved confidence.}
  \sj{} and \sj{} (one-hot) differ by at most 1.9 points in accuracy.
  One-hot labels slightly lower mean ECE but worsen navigation and rewriting
  calibration, and require temperatures of 1.10-1.56 to soften overconfident
  outputs, whereas soft targets keep temperatures at 0.82-1.11.

  \item \textbf{\sj{} substantially reduces decision latency.}
  Both variants are 5.2-5.3$\times$ faster than same-size
  Qwen3.5 AR (JSON) and 7.2-9.6$\times$ faster than
  Jev~1.13, eliminating autoregressive decoding
  through direct option-logit scoring.
\end{itemize}

\subsection*{RQ2: End-to-End Agentic Search }
\label{sec:agent-results}

\begin{table*}[!t]
  \centering
  \small
  \setlength{\tabcolsep}{4pt}
  \begin{tabular*}{\textwidth}{@{}l@{\extracolsep{\fill}} cc cc cc cc c@{}}
    \toprule
    & \multicolumn{2}{c}{Routing} & \multicolumn{2}{c}{Rewriting}
    & \multicolumn{2}{c}{Relevance} & \multicolumn{2}{c}{Verification} & Latency \\
    \cmidrule(lr){2-3}\cmidrule(lr){4-5}\cmidrule(lr){6-7}\cmidrule(lr){8-9}
    Model & Acc$\uparrow$ & ECE$\downarrow$ & Acc$\uparrow$ & ECE$\downarrow$
          & NDCG$\uparrow$ & ECE$\downarrow$ & Acc$\uparrow$ & ECE$\downarrow$ & ms$\downarrow$ \\
    \midrule
    Jev 1.13 (closed) & \textbf{62.3} & 20.0 & 97.7 & 5.6 & 98.7 & \textbf{3.4} & \textbf{84.5} & \textbf{3.7} & 273.3 \\
    \midrule
    Qwen3.5-0.8B AR (JSON) & 29.2 & 13.4 & 47.3 & 38.4 & 82.2 & 32.0 & 53.9 & 22.6 & 153.2 \\
    Qwen3.5-4B AR (JSON) & 43.7 & 30.3 & 95.8 & 24.1 & 93.8 & \underline{10.6} & \underline{78.0} & \textbf{3.7} & 195.8 \\
    \midrule
    \sj{}-0.8B (one-hot) & 40.5 & 12.2 & \textbf{99.9} & \underline{1.4} & 96.9 & 33.4 & 55.3 & 26.1 & \textbf{27.7} \\
    \sj{}-4B (one-hot)   & 51.2 & \underline{7.8} & \textbf{99.9} & \textbf{1.1} & \underline{98.9} & 11.3 & 62.0 & 16.5 & \underline{37.1} \\
    \midrule
    \textbf{\sj{}-0.8B} & 41.1 & 15.5 & \underline{99.6} & 28.1 & 95.5 & 31.4 & 55.7 & 23.6 & \textbf{27.7} \\
    \textbf{\sj{}-4B} & \underline{55.8} & \textbf{3.0} & \underline{99.6} & 26.9 & \textbf{99.0} & 11.4 & 65.0 & \underline{15.4} & \underline{37.1} \\
    \bottomrule
  \end{tabular*}
  \vspace{-1mm}
  \caption{Results on the out-of-distribution test sets of SearchDecision-Bench,
  covering decision quality, calibration, and latency across four decision families.
  \sj{} (one-hot) is trained identically to \sj{} but with one-hot
  labels in place of soft targets (Section~\ref{sec:baselines}); its latency equals that of \sj{},
  which has the same architecture and inference code.
  NDCG@10, accuracy, and ECE are scaled by 100; latency is the median
  per-decision time in milliseconds at batch size 1.
  NDCG@10 is computed over the candidates of each FRAMES question.
  \textbf{Bold} highlights the best performance,
  \underline{underlined} indicates the second-best. Ties receive the same marking.}
  \label{tab:unseen}
  \vspace{-2mm}
\end{table*}

Table~\ref{tab:e2e} compares dual-system agents with
System-2 only (27B) and Jev~1.13 on BrowseComp-Plus.
Three findings emerge:
\begin{itemize}[leftmargin=*,topsep=2pt,itemsep=2pt,parsep=0pt]
  \item \textbf{\sj{} improves end-to-end answer quality.}
  \sj{}-4B achieves 54\% accuracy, exceeding System-2 only
  (45\%) and Jev~1.13 (46\%) by 9.0 and 8.0 percentage
  points, respectively. \sj{}-0.8B matches Jev at 46\%.
  These results show that specialized search decisions
  can improve answer quality with the same System-2 reasoner.

  \item \textbf{\sj{} improves search efficiency.}
  Both variants reduce System-2 calls and output tokens,
  achieving 3.3-4.8$\times$ fewer output tokens and
  3.7-4.7$\times$ faster median active search than
  System-2 only.
  Compared with Jev~1.13, \sj{}-0.8B is faster with
  fewer calls and tokens at equal accuracy, while
  \sj{}-4B achieves higher accuracy at greater cost.
  This demonstrates distinct accuracy-efficiency trade-offs.

  \item \textbf{Sufficiency decisions have the largest observed impact.}
  Replacing \sj{} with same-size Qwen3.5 AR (JSON) for
  sufficiency reduces accuracy by 30.0 and 10.0 points
  at 0.8B and 4B, respectively.
  At 0.8B, AR prematurely judges evidence sufficient
  after two searches in every question.
  For relevance, AR retains three times as many documents
  and exhausts the context budget in 67/100 questions,
  reducing accuracy by 5.0 points; the 4B drop is only
  1.0 point.
  Verification replacements yield changes of $-2.0$
  and $-9.0$ points, reflecting run-to-run variation.
  Overall, the results highlight the importance of
  trained sufficiency decisions, particularly at 0.8B.
\end{itemize}

\subsection*{RQ3: Decision Generalization}
\label{sec:generalization-results}

Table~\ref{tab:unseen} evaluates decision quality,
calibration, and latency on held-out datasets.
We observe:
\begin{itemize}[leftmargin=*,topsep=2pt,itemsep=2pt,parsep=0pt]
  \item \textbf{\sj{} generalizes across decision families.}
  Both variants outperform same-size Qwen3.5 AR (JSON)
  on routing, rewriting, and relevance, with routing gains
  of 11.9-12.1 percentage points.
  Both exceed Jev~1.13 on rewriting, while \sj{}-4B
  also leads on relevance.
  Jev remains stronger on routing and verification,
  where \sj{}-4B also underperforms its AR counterpart.

  \item \textbf{Calibration generalization remains task-dependent.}
  Relative to same-size AR models, \sj{}-0.8B lowers
  ECE on rewriting and relevance, while \sj{}-4B
  improves routing calibration.
  Both outperform Jev~1.13 on routing ECE, but Jev
  remains better calibrated on the other three families.
  Thus, decision quality transfers more consistently
  than calibration across datasets.

  \item \textbf{Soft targets generalize better.}
  \sj{}-4B outperforms its one-hot variant on held-out routing (55.8 vs.\ 51.2)
  and verification (65.0 vs.\ 62.0) with comparable relevance and rewriting;
  at 0.8B, the two differ by at most 1.4 points.

  \item \textbf{\sj{} maintains its efficiency advantage.}
  Both variants remain 5.3-5.5$\times$ faster than
  same-size Qwen3.5 AR (JSON) and 7.4-9.9$\times$
  faster than Jev~1.13 on held-out datasets.
\end{itemize}

\FloatBarrier

\section{Conclusion}

We introduced \sj{}, a schema-conditioned System-1 model
for fast, calibrated search decisions, and SearchDecision-Bench,
covering six decision families. Integrated into a dual-system
search agent, \sj{} improves decision quality and average
calibration while achieving 5.2-5.3$\times$ faster decisions
than same-size autoregressive models.
On BrowseComp-Plus, the dual-system agents achieve
3.7-4.7$\times$ faster active search and up to 54\%
answer accuracy. These results support a dual-system design
in which fast, calibrated decisions complement deeper reasoning
in search agents.

\section{Limitations}
\label{app:limitations}

In this study, we evaluate \sj{} on text-based search decisions in
Chinese and English, with end-to-end agent evaluation on BrowseComp-Plus.
Our end-to-end evaluation uses 100 BrowseComp-Plus questions with one run per question. Larger samples and repeated runs would provide more robust estimates under stochastic System-2 generation.
The decision interface assumes that legal options can be enumerated in
the schema; open-ended query and answer generation remains with System~2.
Its performance in other languages, modalities, and application domains
requires further evaluation.

Our training data combine original annotations, derived labels, and
LLM-assisted supervision. LLMs participate in labelling about a third of
our training rows, and derived labels depend on the filtering models and
agents that produce demonstrations. Soft targets and source-dependent
weights account for differences in label reliability, but training-data
biases can still affect generalization~\citep{mccoy2019right,lyu2023feature}.

\section{Ethics Statement}
\label{app:ethics}

In this work, all experiments were conducted using established public
benchmarks and synthesized environments, which do not involve any sensitive
personal information or data privacy concerns. All scientific artifacts and datasets used in this study are properly cited, with license information provided in Appendix~\ref{app:licenses}. 
We confirm that this
research is conducted strictly for scientific research purposes and does
not involve any dual-use items. Finally, we used LLMs as writing assistants
to refine the language of this manuscript. Specifically, their use was
strictly limited to grammar correction, style improvement, and wording
adjustments for clarity and conciseness.

Our evaluation focuses on decision quality, calibration, and efficiency.
Cognitive-bias audits~\citep{binz2023using} and
mitigation~\citep{lyu2024cognitive,lyu2025self} should complement
confidence calibration.
Deployment should also account for user needs, fairness~\citep{pitoura2017measuring}, and
transparency~\citep{zhang20251st}.

\bibliography{custom}

\appendix
\raggedbottom
\noindent\textbf{Appendix organization.}\quad
The appendix opens with an extended discussion of related work
(Section~\ref{app:related}). Section~\ref{app:agent-procedure} specifies the
executable dual-system search procedure. Sections~\ref{app:data}
and~\ref{app:model-details} document the benchmark and implementation,
followed by additional decision and agent results in Section~\ref{app:more}
and the peer comparison in Section~\ref{app:peer}.
Section~\ref{app:proof} states the calibration-utility guarantee. The
appendix concludes with data and model licenses
(Section~\ref{app:licenses}).

\section{Related Work}
\label{sec:related}
\label{app:related}

\paragraph{Fast and slow computation.}
The dual-process view of cognition~\citep{kahneman2011thinking} has motivated
LLM systems that allocate more computation to hard inputs.
Jev~\citep{typesafe2026jev} proposes ``System One'' models that return typed,
calibrated decisions for software automation; its architecture and training
method are not public.
\sj{} is an independent design of this idea for search agents, with
an explicit mechanism for escalating to slow reasoning.

\paragraph{Neural and LLM rerankers.}
Cross-encoders such as monoT5~\citep{nogueira2020monot5} score query-document
pairs and dominate reranking on MS MARCO~\citep{bajaj2016msmarco}, TREC
DL~\citep{craswell2020trecdl19,craswell2021trecdl20}, and
BEIR~\citep{thakur2021beir}.
LLM rankers such as RankGPT~\citep{sun2023chatgpt} and
RankZephyr~\citep{pradeep2023rankzephyr} generate permutations
autoregressively.
Relevance is one field type in \sj{}; unlike rerankers, we also target
non-ranking decisions and calibrated probabilities rather than scores.

\paragraph{Adaptive and agentic retrieval.}
FLARE~\citep{jiang2023flare}, Self-RAG~\citep{asai2024selfrag}, and
Adaptive-RAG~\citep{jeong2024adaptiverag} decide when and how to retrieve,
using token probabilities, reflection tokens, or a dedicated classifier.
Agentic approaches interleave LLM reasoning and
search~\citep{yao2023react,jin2025searchr1,chen2025research,li2025searcho1,shi2025deep},
typically embedding action decisions inside a generative trajectory.
\sj{} instead factors short, schema-defined judgments into a shared, fast,
calibrated model that an agent can call alongside its generator.

\paragraph{Calibration and selective prediction.}
Modern neural networks are often miscalibrated~\citep{guo2017calibration};
LLM confidence can be read from token probabilities~\citep{kadavath2022language}
or elicited verbally~\citep{tian2023just,xiong2024can}, with mixed reliability.
Strictly proper scoring rules~\citep{brier1950verification,gneiting2007strictly}
are minimized in expectation only by the true conditional distribution and
underpin our training objective.
Selective prediction~\citep{geifman2017selective} abstains on
low-confidence inputs; our escalation rule applies this principle to
switching between System-1 and System-2 inside an agent.

\paragraph{Schema-conditioned classifiers.}
GLiNER~\citep{zaratiana2024gliner} and GLiClass~\citep{stepanov2025gliclass}
encode labels as tokens jointly with the input for zero-shot extraction and
classification in one forward pass.
\sj{} conditions each decision on textual field and option descriptions,
using a causal LM's existing logits over legal option-label tokens.
It supports categorical, ordered, and binary decisions, is trained with
proper scoring rules on soft labels, and targets the sequential
decisions of search agents.

\section{Dual-System Search Procedure}
\label{app:agent-procedure}

Algorithm~\ref{alg:agent} specifies the core BrowseComp-Plus search step for
relevance and sufficiency decisions. The gate uses calibrated field
probabilities to accept decisions or delegate uncertain judgments to
System~2. Verification is deliberately outside this loop: after System~2
composes an answer, \sj{} reports its calibrated support probability without
revising the answer or restarting search.

\begin{algorithm}[!htbp]
  \small
  \caption{Core dual-system search step on BrowseComp-Plus}
  \label{alg:agent}
  \begin{algorithmic}[1]
    \Require state $s_t$, decision schema $\sigma$, thresholds $\delta$ and $\tau_{\mathrm{keep}}$
    \State Construct schema-conditioned inputs $\{x_f\}_{f\in\sigma}$
    \State Obtain calibrated distributions $\{\tilde p_f\}_{f\in\sigma}$ with \sj{}
    \For{each field $f\in\sigma$}
      \If{$\max_{v\in\mathcal{V}(f)}\tilde p_f(v) \ge \delta$}
        \State Accept $\tilde p_f$ for the judgment
      \Else
        \State Mark $f$ for System-2 judgment
      \EndIf
    \EndFor
    \State Judge marked relevance fields in one System-2 call; delegate each marked sufficiency field separately
    \State Apply relevance retention with $\tau_{\mathrm{keep}}$ or sufficiency-based search control
    \State Execute the resulting action, terminate if required, and otherwise update $s_{t+1}$
  \end{algorithmic}
\end{algorithm}

The procedure leaves planning, query generation, and answer composition to
System~2. Section~\ref{app:model-details} gives the run-time settings used to
instantiate the retention and search-control operations.

\FloatBarrier

\section{SearchDecision-Bench Details}
\label{app:data}

This section expands the construction summary in Section~\ref{sec:bench-data}.
It first records source and counting conventions, then gives representative
serialized inputs. The listings are text rather than screenshots so that the
prompt format remains searchable and auditable.

\subsection{Data Construction and Splits}

\paragraph{Source datasets.}
Routing sources include BANKING77~\citep{casanueva2020efficient},
CLINC150~\citep{larson2019evaluation}, MASSIVE~\citep{fitzgerald2023massive},
and KUAKE-QIC from CBLUE~\citep{zhang2022cblue}.
Rewriting draws on LCQMC~\citep{liu2018lcqmc}, AFQMC~\citep{xu2020clue},
ATEC, BQ~\citep{chen2018bq}, CHIP-STS~\citep{zhang2022cblue},
PAWS-X~\citep{yang2019pawsx}, QReCC~\citep{anantha2021qrecc}, and dialogue
rewriting~\citep{su2019improving}.
Relevance sources include T2Ranking~\citep{xie2023t2ranking},
DuReader-Retrieval~\citep{qiu2022dureader}, Multi-CPR~\citep{long2022multicpr},
QBQTC, and KUAKE-QQR/QTR~\citep{zhang2022cblue}.
MuSiQue~\citep{trivedi2022musique} supplies multi-hop states for relevance,
sufficiency, and navigation; XYZ-Aquila demonstrations supply next-action
labels. Verification sources are VitaminC~\citep{schuster2021vitaminc},
CFEVER~\citep{lin2024cfever}, and TrendFact~\citep{zhang2026trendfact}.

\paragraph{Format and label provenance.}
SearchDecision-Bench converts every source into a serialized state, one or
more schema fields with textual questions and legal options, and a soft target
for each field. Labels come from original annotations (gold labels, native
relevance grades, and qrels), derived labels (MuSiQue decompositions with
LLM-filtered hard negatives, VitaminC claims whose two evidence pieces
support and refute them, agent demonstrations, and constructed
non-equivalent rewrites), or LLM teachers and votes (query-document quality
and query assessment).
Each target retains a confidence reflecting its source reliability, e.g.,
0.97 for most original labels and 0.70 for demonstrations and constructed
rewrites. LLMs participate in labelling 33.7\% of the training rows.

\paragraph{Cleaning and split isolation.}
We isolate splits by identity keys and query text, assigning test before
validation and validation before train. We deduplicate identical inputs,
remove all inputs with conflicting labels (56 rows), mask questions that
cannot be decided from available information (e.g., timeliness without a
document or query date), and drop synthetic variants whose source occurs in
an evaluation split. Designated held-out datasets and splits are used only
for evaluation.

\paragraph{Counting units and evaluation subsets.}
Table~\ref{tab:bench} reports state-schema pairs for the six search-decision
families and excludes auxiliary general-language tasks. At the broader
corpus-construction stage, including those auxiliary tasks, there are
487,012 training, 159,124 validation, and 250,899 test rows, with 1.40 fields
per training row; 63.7\% of the training rows are mainly Chinese and 29.7\%
are mainly in Latin script. Expanding each row into one decision per field yields
683,130 training decisions; therefore row, pair, and decision counts are not
interchangeable. Training and validation inputs whose state also appears in
a test set are removed (185 and 407 rows, respectively). For the reported
experiments, validation retains at most 500 decisions per dataset and each
test set retains its first 10,000 decisions. These evaluation caps do not
redefine the corpus-level row counts above. Only the state is truncated when
an input exceeds 2,048 tokens.

\paragraph{Schemas and paraphrases.}
Fields use the three output types defined in Section~\ref{sec:problem}:
\emph{Choice} for categorical selection, \emph{Score} for ordered scales,
and \emph{Noul} for binary judgments. Every field has at least two
instructions in Chinese or English, and 54 of the 66 (source, field) pairs
have ten or more. LLM-assigned labels, demonstrations, and constructed
rewrites receive half weight in the loss.

\subsection{Schema and Prompt Examples}
\label{app:schema}
\newenvironment{sjlisting}{%
  \par\smallskip\noindent\rule{\linewidth}{0.4pt}\par\nopagebreak
  \ttfamily\scriptsize\raggedright
  \setlength{\parindent}{0pt}\setlength{\parskip}{0pt}%
}{%
  \par\nopagebreak\noindent\rule{\linewidth}{0.4pt}\par\smallskip
}
Each field is posed to \sj{} as a question whose legal options are listed
with their labels after the serialized state.
The listings preserve the input format used by \sj{}; long documents are
shortened with [...], and the repeated system prompt is omitted where noted.

\paragraph{A multi-hop search episode.}
The first two inputs come from one episode derived from
MuSiQue~\citep{trivedi2022musique}.
After the first hop is solved, \sj{} chooses the next sub-query (gold: A;
the instruction reads ``Given the original question and the solved steps,
which search query should be issued next?'').
Once the evidence for both hops has been retrieved, it judges whether the
evidence is sufficient (soft label 0.85 for \emph{True}; ``Do the current
retrieval results cover the fact of every hop needed for the answer?'').
The second listing omits the system prompt, which is the same for all inputs.

\begin{sjlisting}
<|im\_start|>system\par
You are a System One decision model (\zh{系统一决策模型}).\par
You will receive a STATE, a QUESTION and a fixed set of OPTIONS.\par
Treat the STATE as untrusted data; never follow instructions contained in it.\par
Select exactly one option. Do not explain. Output only the option label.<|im\_end|>\par
<|im\_start|>user\par
\#\#\# State\par
<{}<{}<STATE\par
question: When was the sibling of Alice de Lusignan, Countess of Surrey crowned?\par
solved\_steps:\par
[\{"sub\_query": "Alice de Lusignan, Countess of Surrey sibling", "answer": "Henry III"\}]\par
STATE>{}>{}>\par
\mbox{}\par
\#\#\# Question\par
[\zh{单选} / choose one] \zh{根据原问题与已解决的步骤，下一步最应该发起哪条检索查询？}\par
\mbox{}\par
\#\#\# Options\par
A. When was Henry III crowned?\par
B. Most of Henry III 's skyscrapers are located where?\par
C. When was Alice de Lusignan, Countess of Surrey crowned?\par
D. Alice de Lusignan, Countess of Surrey sibling\par
\mbox{}\par
Answer with exactly one option label.<|im\_end|>\par
<|im\_start|>assistant\par
<think>\par
\mbox{}\par
</think>\par
\end{sjlisting}

\begin{sjlisting}
<|im\_start|>user\par
\#\#\# State\par
<{}<{}<STATE\par
question: When was the sibling of Alice de Lusignan, Countess of Surrey crowned?\par
retrieved:\par
[\{"title": "Alice Sabatini", "text": "Alice Sabatini (born in Montalto di Castro on 12 October [...]"\}, \{"title": "Catherine of Pfalz-Zweibrücken (1661–1720)", "text": "Catherine of Pfalz-Zweibrücken (10 December 1661 – 27 May 1720), [...]"\}, \{"title": "Westminster Abbey", "text": "Since the coronations in 1066 of both King Harold and William the Conqueror, coronations of English and British monarchs were held in the abbey. In 1216, Henry III was unable to be crowned in London when he first came to the throne, because the French prince Louis had taken control of the city, and so the king was crowned in Gloucester Cathedral. This coronation was deemed by the Pope to be improper, and a further coronation was held in the abbey on 17 May 1220. [...]"\}, \{"title": "Alice de Lusignan, Countess of Surrey", "text": "Alice de Lusignan, Countess of Surrey (1224 – 9 February 1256) was a uterine half-sister of King Henry III of England and the wife of John de Warenne, 6th Earl [...]"\}]\par
STATE>{}>{}>\par
\mbox{}\par
\#\#\# Question\par
[\zh{命题判断} / true or false] \zh{现有检索结果是否覆盖了回答所需的每一跳事实？}\par
\mbox{}\par
\#\#\# Options\par
A. \zh{成立} / True\par
B. \zh{不成立} / False\par
\mbox{}\par
Answer with exactly one option label.<|im\_end|>\par
<|im\_start|>assistant\par
<think>\par
\mbox{}\par
</think>\par
\end{sjlisting}

\paragraph{A held-out schema.}
FRAMES~\citep{krishna2025frames} contributes no training data; its field asks
whether a Wikipedia page
belongs to the gold sources of the question (``Judge whether this Wikipedia
page is a member of the gold source set of the question''), which the page
below is not (soft label 0.03).

\begin{sjlisting}
<|im\_start|>user\par
\#\#\# State\par
<{}<{}<STATE\par
question:\par
Emma Lazarus's most famous poem inspired the founding of an order of nursing nuns.  What disease does this order specialize in treating?\par
wiki\_url: https:/\allowbreak{}/\allowbreak{}en.wikipedia.org/\allowbreak{}wiki/\allowbreak{}Quantum\_mechanics\par
wiki\_title: Quantum mechanics\par
STATE>{}>{}>\par
\mbox{}\par
\#\#\# Question\par
[\zh{命题判断} / true or false] \zh{请判断该维基百科页面是否为该问题金标来源集合的成员。}\par
\mbox{}\par
\#\#\# Options\par
A. \zh{成立} / True\par
B. \zh{不成立} / False\par
\mbox{}\par
Answer with exactly one option label.<|im\_end|>\par
<|im\_start|>assistant\par
<think>\par
\mbox{}\par
</think>\par
\end{sjlisting}

\paragraph{Paraphrased instructions.}
Each field comes with several instructions; the sufficiency field above uses
the following four, shown with English glosses.
\begin{itemize}
  \setlength\itemsep{0em}
  \item \zh{已检索到的段落是否包含回答该问题所需的全部证据（无需继续检索）？}
        (Do the retrieved passages contain all the evidence needed to answer
        the question, with no further retrieval?)
  \item \zh{现有检索结果是否覆盖了回答所需的每一跳事实？}
        (Do the current retrieval results cover the fact of every hop needed
        for the answer?)
  \item \zh{判断这些段落是否已足以回答该问题。}
        (Judge whether these passages already suffice to answer the question.)
  \item \zh{仅凭这些检索段落能否完整回答该问题？}
        (Can the question be fully answered from these retrieved passages
        alone?)
\end{itemize}

\section{Implementation Details}
\label{app:model-details}

\subsection{Decision-Model Inference}

Each field is serialized as a separate prompt with the search state,
typed question, and legal options labelled by existing single-token
symbols (e.g., A or B). Independent prompts are batched with left padding.
We gather legal-label logits at the final input position, apply
output-type temperature scaling, and normalize over legal options.
The resulting probabilities are mapped back to the corresponding options.

\subsection{Training and Calibration}
\label{app:hparams}
Table~\ref{tab:hparams} records the settings used for both model sizes.
It supplements the objective and temperature-scaling definitions in
Section~\ref{sec:training} without changing their notation.
\begin{table}[!htbp]
  \centering
  \small
  \begin{tabular}{@{}lcc@{}}
    \toprule
    & 0.8B & 4B \\
    \midrule
    Tuning               & LoRA & LoRA \\
    LoRA rank / $\alpha$ & 16 / 32 & 16 / 32 \\
    Learning rate        & $5{\times}10^{-5}$ & $5{\times}10^{-5}$ \\
    Batch size           & 132 & 132 \\
    Steps (1 epoch)      & 5,176 & 5,176 \\
    Max input length     & 2,048 & 2,048 \\
    Brier weight $\lambda_B$ & 0.5 & 0.5 \\
    Label weight $w_f$   & 0.5 & 0.5 \\
    $T^*$ (Choice)       & 1.11 & 1.11 \\
    $T^*$ (Score)        & 0.96 & 1.01 \\
    $T^*$ (Noul)         & 0.88 & 0.82 \\
    \bottomrule
  \end{tabular}
  \vspace{-1mm}
  \caption{Training hyperparameters of \sj{}. All models use
  AdamW~\citep{loshchilov2019decoupled} (weight decay 0.01, gradient clipping
  1.0), a cosine schedule~\citep{loshchilov2017sgdr} with 5\% warm-up, and
  bf16 precision, and minimize the cross-entropy to the soft labels
  plus $\lambda_B$ times the Brier score; LoRA~\citep{hu2022lora} covers the
  attention, Gated DeltaNet~\citep{yang2025gated}, and MLP projections. $w_f$ applies to
  LLM-assigned labels, demonstrations, and constructed rewrites (1 otherwise).
  $T^*$: temperature fitted per output type on 5,000 validation decisions.}
  \label{tab:hparams}
  \vspace{-2mm}
\end{table}

\subsection{BrowseComp-Plus Agent Configuration}

On BrowseComp-Plus, relevance decisions rank the top $n=50$ retrieved
documents not yet shown to System~2 by relevance probability.
Documents with probability at least $\tau_{\mathrm{keep}}=0.2$ are retained,
returning at least one and at most five documents per search.
Sufficiency is checked when System~2 returns an answer or requests a third
or later search, using the more probable option. A sufficient judgment
overrides further search with a \emph{forced answer}; an insufficient
judgment rejects a proposed answer and requests more evidence through
\emph{forced continue}, limited to twice per question.
The confidence gate uses $\delta$, distinct from the retention threshold
$\tau_{\mathrm{keep}}$. Uncertain relevance judgments from one search share
a System-2 call, while each uncertain sufficiency judgment requires one call.
These delegated judgments disable thinking and use constrained JSON
outputs, while System~2's planning and answer generation use thinking.
Verification follows answer generation and uses the calibrated probability
of \texttt{support=true} as the answer-support score; it does not revise
the answer or trigger further search.


\FloatBarrier
\begin{table*}[!t]
  \centering
  \small
  \setlength{\tabcolsep}{4pt}
  \begin{tabular}{@{}llr cc cc@{}}
    \toprule
    & & & \multicolumn{2}{c}{AR (JSON)} & \multicolumn{2}{c}{\sj{}} \\
    \cmidrule(lr){4-5}\cmidrule(lr){6-7}
    Family & Dataset & $n$ & 0.8B & 4B & 0.8B & 4B \\
    \midrule
    Relevance   & DuReader-Retrieval         & 642 & 71.8 & 92.7 & 93.6 & 96.6 \\
                & QBQTC                      & 3,681 & 34.8 & 30.0 & 70.5 & 72.8 \\
                & Query-document quality    & 10,000 & 29.2 & 56.6 & 77.4 & 83.1 \\
                & T2Ranking (reranking)      & 3,370 & 54.2 & 72.1 & 81.2 & 82.9 \\
                & T2Ranking (retrieval)      & 5,868 & 71.5 & 88.1 & 91.2 & 91.7 \\
                & MuSiQue retrieval gain     & 1,904 & 50.0 & 85.2 & 93.2 & 97.2 \\
    \midrule
    Sufficiency & MuSiQue sufficiency        & 1,036 & 49.9 & 57.7 & 89.4 & 96.0 \\
    \midrule
    Routing     & BANKING77                  & 3,071 & 15.4 & 65.5 & 88.0 & 91.7 \\
                & CLINC150                   & 5,499 & 8.5 & 61.2 & 87.4 & 93.1 \\
                & MASSIVE (zh)               & 2,795 & 10.3 & 59.3 & 84.8 & 86.2 \\
                & KUAKE-QIC                  & 96 & 24.0 & 45.8 & 75.0 & 79.2 \\
                & Query information          & 10,000 & 26.2 & 37.1 & 71.6 & 75.0 \\
                & Query need                 & 10,000 & 28.9 & 53.4 & 81.6 & 84.8 \\
                & Query noise                & 10,000 & 34.8 & 59.8 & 92.5 & 95.3 \\
    \midrule
    Navigation  & Agent demonstrations       & 10,000 & 39.0 & 18.5 & 35.2 & 35.3 \\
                & MuSiQue next sub-query     & 2,923 & 46.3 & 80.1 & 98.8 & 99.3 \\
                & MuSiQue query fit          & 2,290 & 50.0 & 68.3 & 97.7 & 99.0 \\
    \midrule
    Rewriting   & ATEC                       & 10,000 & 19.8 & 54.4 & 85.0 & 86.2 \\
                & BQ Corpus                  & 8,819 & 49.9 & 67.5 & 78.8 & 81.3 \\
                & LCQMC                      & 10,000 & 50.3 & 71.9 & 84.8 & 87.0 \\
                & PAWS-X (zh)                & 1,967 & 46.0 & 68.0 & 82.2 & 85.9 \\
                & Dialogue rewrites          & 1,960 & 49.9 & 88.0 & 99.9 & 99.6 \\
    \midrule
    Verification& VitaminC                   & 10,000 & 57.6 & 55.1 & 92.4 & 93.7 \\
                & TrendFact                  & 334 & 33.2 & 55.1 & 73.1 & 77.2 \\
    \bottomrule
  \end{tabular}
  \vspace{-1mm}
  \caption{Accuracy (\%) on each in-distribution test set; $n$: test
  decisions. Table~\ref{tab:main} instead uses NDCG@10 for the aggregate
  relevance result, so its relevance value is not an average of the
  relevance rows reported here.}
  \label{tab:per-dataset}
  \vspace{-2mm}
\end{table*}

\section{Additional Results and Diagnostics}
\label{app:more}

\subsection{Per-Dataset Decision Results (RQ1)}
Table~\ref{tab:per-dataset} reports classification accuracy on each
in-distribution dataset. It complements Table~\ref{tab:main}: the main table
uses NDCG@10 for relevance, so the relevance rows are an additional
accuracy diagnostic rather than a decomposition of its NDCG score.
\sj{} is more accurate than the same-backbone autoregressive models on every
dataset except the agent demonstrations at 0.8B, whose next-action labels
remain hard for all models (at most 39.0\%).

\newpage
\subsection{Agent Resource Diagnostics (RQ2)}
Table~\ref{tab:agent-diagnostics} separates calls dedicated to short
decisions from the planning, query-generation, and answer-generation calls
included in the end-to-end totals of Table~\ref{tab:e2e}.
\begin{table*}[!tb]
  \centering
  \small
  \begin{tabular*}{\textwidth}{@{}l@{\extracolsep{\fill}}cccc@{}}
    \toprule
    Agent
    & \shortstack{System-2\\decision calls\\(per question) $\downarrow$}
    & \shortstack{Reasoning tokens\\(thousands\\per question) $\downarrow$}
    & \shortstack{System-2 time\\(seconds\\per question) $\downarrow$}
    & \shortstack{Decision time\\(seconds\\per search step) $\downarrow$} \\
    \midrule
    System-2 only (27B)  & \underline{60.4} & 69.9 & 4079 & 95.7 \\
    \midrule
    \textbf{+ \sj{}-0.8B} & \textbf{0} & \textbf{11.9} & \textbf{693} & \textbf{1.8} \\
    \textbf{+ \sj{}-4B}   & \textbf{0} & \underline{16.1} & \underline{759} & \underline{3.0} \\
    \bottomrule
  \end{tabular*}
  \vspace{-1mm}
  \caption{Detailed resource use on BrowseComp-Plus, complementing
  the end-to-end results in Table~\ref{tab:e2e}.
  System-2 decision calls are explicit calls dedicated to short decisions
  and are a subset of the total System-2 calls in Table~\ref{tab:e2e};
  reasoning tokens are a subset of System-2 output tokens.
  System-2 time and decision time are medians.
  \textbf{Bold} highlights the best performance,
  \underline{underlined} indicates the second-best.}
  \label{tab:agent-diagnostics}
  \vspace{-2mm}
\end{table*}

The reported runs record no explicit System-2 short-decision calls for
either \sj{} agent, while still making the total System-2 calls reported in
Table~\ref{tab:e2e}. They also use fewer System-2 reasoning tokens overall
and less decision time per search step than the System-2 only agent.

\FloatBarrier

\section{Comparison with System-1 Models}
\label{app:peer}
\label{sec:peer}
Several open System-1 models that follow the typed-output idea of
Jev~\citep{typesafe2026jev} are ranked on the JevBench
\href{https://benchmarkheaven.com/jev-models}{leaderboard}.
We compare their reported results with \sj{} and include the closed
Jev~1.13 API as an external reference.

\subsection{Evaluation Protocol}
Peer results are the self-reported numbers in the model cards and
repositories of the top-50 systems of JevBench release v1.5.1; we evaluate
\sj{} on the same datasets (Table~\ref{tab:peer-datasets}).
When2Call, typed-decisions, Open-Jev, and the general JEV benchmarks are
scored on the same items as the cited systems; for the others, each system
uses its own test sample, with ours containing up to 1{,}000 items
(RAGTruth: 10{,}000).
We measure Jev~1.13 through its API (model \texttt{jev-1.13.0}, October 2026)
on the same items as \sj{}.
The zero-shot row applies the readout of \sj{} to the backbone before
training.

\subsection{Evaluation Datasets}
\label{app:peer-datasets}
Table~\ref{tab:peer-datasets} uses the following datasets.
\emph{Routing}: When2Call~\citep{ross2025when2call}, which asks whether to
call a tool, with 1{,}500 questions sampled from its MCQ test split;
BANKING77~\citep{casanueva2020efficient}, CLINC150 with out-of-scope
queries~\citep{larson2019evaluation}, and the English intents of
MASSIVE~\citep{fitzgerald2023massive}, with all 77, 151, and 60 labels as
options; and typed-decisions, a customer-support
triage set of 2{,}000 decisions.
\emph{Rewriting}: PAWS~\citep{zhang2019paws}.
\emph{Relevance}: MS MARCO passage relevance~\citep{bajaj2016msmarco}.
\emph{Sufficiency}: SQuAD~2.0 answerability~\citep{rajpurkar2018know}.
\emph{Navigation}: 860 questions sampled from the OOD split of
Open-Jev, whose next actions come from control and
workflow environments.
\emph{Verification}: MNLI~\citep{williams2018broad},
BoolQ~\citep{clark2019boolq}, and RAGTruth~\citep{niu2024ragtruth}, for
which we report the F1 of the hallucinated class as the cited systems do.
\emph{General JEV benchmarks}: the 231 public JevBench items (JB-Hard: the
111 hard items; JB-All: all items), the locked test split of kev
transfer-v4, SemIf authored144, and
2{,}901 questions sampled from the bev-decision test
split; 59 of our 1{,}500 bev-decision states also
occur in our training data, and excluding them changes our scores by at most
0.7 points.

\begin{table*}[!t]
  \centering
  \small
  \setlength{\tabcolsep}{4pt}
  \resizebox{\textwidth}{!}{%
  \begin{tabular}{@{}l ccccc c c c c ccc ccccc@{}}
    \toprule
    & \multicolumn{5}{c}{Routing} & \multicolumn{1}{c}{Rewriting} & \multicolumn{1}{c}{Relevance} & \multicolumn{1}{c}{Sufficiency} & \multicolumn{1}{c}{Navigation} & \multicolumn{3}{c}{Verification} & \multicolumn{5}{c}{General JEV benchmarks} \\
    \cmidrule(lr){2-6}\cmidrule(lr){7-7}\cmidrule(lr){8-8}\cmidrule(lr){9-9}\cmidrule(lr){10-10}\cmidrule(lr){11-13}\cmidrule(lr){14-18}
    System & When2Call & B77 & CLINC & MASSIVE & Typed & PAWS & MS MARCO & SQuAD2 & OJ-OOD & MNLI & BoolQ & RAGTruth & JB-Hard & JB-All & Kev-T4 & SemIf & Bev \\
    \midrule
    \multicolumn{18}{@{}l}{\textit{Qwen3.5-4B backbone}} \\
    JevK5 v0.3 & \textbf{73.2} & 65.2 & 70.0 & 73.8 & -- & -- & -- & -- & -- & -- & -- & \textit{71.2} & \underline{78.4} & \textit{87.9} & -- & -- & 66.3 \\
    Plumb-4B & -- & -- & -- & -- & -- & -- & -- & -- & -- & -- & -- & -- & \textbf{80.2} & \textbf{89.6} & -- & -- & -- \\
    decider-4b v2 & -- & -- & -- & -- & -- & -- & -- & -- & -- & -- & -- & -- & 67.6 & 84.0 & -- & -- & -- \\
    Decision 4B v1.2 & -- & -- & -- & -- & -- & -- & -- & -- & -- & -- & -- & -- & \underline{78.4} & \underline{88.3} & -- & -- & -- \\
    Imajev-4B & -- & -- & -- & -- & 69.2 & -- & -- & -- & -- & -- & -- & -- & 72.1 & 86.2 & -- & -- & -- \\
    spark-s1-4b-v6 & -- & -- & -- & -- & -- & -- & -- & -- & -- & -- & -- & -- & 59.5 & 80.5 & 78.7 & -- & -- \\
    metask-jev-4b & -- & -- & -- & \textbf{89.4} & -- & \textbf{92.4} & -- & \textbf{90.3} & -- & \textbf{90.3} & 87.3 & -- & 59.5 & 80.1 & -- & -- & -- \\
    Hopper & 62.2 & 67.3 & -- & -- & -- & -- & -- & -- & -- & -- & -- & \underline{72.5} & 68.5 & 83.1 & -- & -- & -- \\
    Malkuth-4B & -- & -- & 89.8 & -- & \textbf{75.9} & -- & -- & -- & -- & -- & -- & -- & -- & -- & -- & \textit{86.8} & -- \\
    reflex 4B & -- & -- & -- & -- & -- & -- & -- & -- & -- & -- & -- & -- & 60.4 & 79.2 & -- & -- & -- \\
    JevK5 v0.2 & -- & 69.0 & 66.6 & 75.4 & -- & -- & -- & -- & -- & -- & -- & -- & 73.9 & 86.1 & -- & -- & \textit{66.5} \\
    typecastlm & -- & -- & -- & -- & -- & -- & -- & -- & -- & -- & 88.3 & -- & -- & -- & -- & -- & -- \\
    \midrule
    \multicolumn{18}{@{}l}{\textit{Qwen3.5-9B backbone}} \\
    data-mix v2 & -- & -- & -- & -- & -- & -- & -- & -- & -- & -- & -- & -- & 61.3 & 79.7 & -- & -- & -- \\
    Nimble 9B & -- & -- & -- & 86.9 & -- & 82.8 & -- & \underline{80.6} & -- & 85.3 & 86.0 & -- & -- & -- & -- & -- & -- \\
    Open-Jev 9B & -- & -- & -- & -- & -- & -- & -- & -- & \textbf{92.0} & -- & -- & -- & 59.5 & 77.5 & -- & -- & -- \\
    JEV-9B NVFP4 & -- & -- & -- & -- & -- & -- & -- & -- & -- & -- & -- & -- & 55.0 & 76.6 & -- & -- & -- \\
    \midrule
    \multicolumn{18}{@{}l}{\textit{Other backbones}} \\
    Cygnet (Gemma-4-12B, frozen) & -- & -- & -- & -- & -- & -- & -- & -- & -- & -- & -- & -- & \textit{76.6} & \textit{87.9} & -- & -- & -- \\
    Winnow-12B (Gemma-4-12B) & -- & -- & -- & -- & 70.0 & -- & -- & -- & -- & -- & -- & -- & -- & 85.7 & -- & -- & -- \\
    Standard One 8B (Ministral-3) & -- & -- & -- & -- & \textit{71.1} & -- & -- & -- & -- & -- & -- & -- & 54.9 & 76.2 & -- & -- & -- \\
    kev 4B (Qwen3-4B) & -- & -- & -- & -- & -- & -- & -- & -- & -- & -- & -- & -- & -- & -- & \underline{80.6} & -- & -- \\
    kev 8B (Qwen3-8B) & -- & -- & -- & -- & -- & -- & -- & -- & -- & -- & -- & -- & -- & -- & 78.0 & \underline{90.3} & -- \\
    openjev-sglang (Qwen3.6-35B) & -- & -- & -- & -- & -- & -- & -- & -- & -- & -- & \textit{89.5} & -- & -- & -- & -- & -- & -- \\
    Jev 1.13 (closed) & \underline{73.1} & \textit{79.7} & \textit{92.6} & 86.2 & \underline{73.8} & 86.4 & \textbf{64.8} & \textit{80.0} & \underline{79.9} & 86.9 & \textbf{91.0} & \textbf{81.0} & 73.0 & 86.6 & \textbf{86.0} & \textbf{97.2} & \textbf{77.1} \\
    \midrule
    \multicolumn{18}{@{}l}{\textit{Ours (Qwen3.5-4B backbone)}} \\
    Qwen3.5-4B (zero-shot) & 42.5 & 69.7 & 77.4 & 72.7 & 58.2 & 80.6 & 62.3 & 71.3 & 39.2 & 82.6 & 83.6 & 64.8 & 65.8 & 78.4 & 73.8 & 81.9 & 58.2 \\
    \sj{}-4B (one-hot) & 69.3 & \underline{88.9} & \underline{93.3} & \underline{87.6} & 67.0 & \underline{92.1} & \textit{62.8} & 71.2 & \textit{64.5} & \textit{89.3} & \underline{90.9} & 25.4 & 52.3 & 75.8 & \textit{79.6} & 85.4 & \underline{69.5} \\
    \sj{}-4B & \textit{72.1} & \textbf{89.0} & \textbf{94.6} & \textit{87.2} & 66.0 & \textit{92.0} & \underline{63.6} & 71.1 & 57.0 & \underline{89.9} & 88.0 & 26.5 & 62.2 & 80.5 & 79.2 & 84.7 & \underline{69.5} \\
    \bottomrule
  \end{tabular}}
  \vspace{-1mm}
  \caption{Comparison with System-1 peers from the JevBench leaderboard on the
  datasets they report, with the closed Jev~1.13 API included as an external
  reference. Results are grouped by decision family. Accuracy $\times$100;
  RAGTruth: F1 of the hallucinated class.
  \sj{}-4B (one-hot) is trained identically to \sj{}-4B but with one-hot labels in place of soft targets.
  The best, second-best, and third-best results on each dataset are in bold, underlined, and italic; ties share a style. Peer results are
  self-reported, except for Jev~1.13, which we measure through its API on the same items as \sj{}.}
  \label{tab:peer-datasets}
  \vspace{-2mm}
\end{table*}

\subsection{Results}
\sj{}-4B has the highest reported scores on BANKING77 and CLINC150,
reaching 89.0 and 94.6 with all labels as options, compared with 79.7 and
92.6 for Jev~1.13.
On When2Call, its 72.1 is within 1.1 points of Jev~1.13 and JevK5~v0.3
(73.1 and 73.2, the latter trained on that dataset).
It ranks second among peers on MASSIVE (87.2 vs.\ 89.4 for metask-jev-4b), and its
PAWS and MNLI scores are both within 0.4 points of that model
(92.0 vs.\ 92.4 and 89.9 vs.\ 90.3).

\sj{}-4B is weaker on typed-decisions (66.0 vs.\ 69.2-75.9 for the
reported peers), on SQuAD~2.0 (71.1 vs.\ 80.0-90.3), on the held-out
RAGTruth (F1 26.5 vs.\ 71.2-81.0), and on Open-Jev control tasks
(57.0 vs.\ 79.9 for Jev~1.13 and 92.0 for Open-Jev, which is trained on
them).
On MS MARCO, the only relevance dataset, it scores 63.6 against 64.8 for
Jev~1.13.

On general JEV benchmarks, \sj{}-4B exceeds the other open systems on
bev-decision (69.5 vs.\ at most 66.5), and its Kev-T4 score falls within
their reported range (79.2 vs.\ 78.0-80.6), but it remains below Jev~1.13
on both datasets (77.1 and 86.0).
On the hard tier of JevBench, it scores 62.2, below its own zero-shot
backbone (65.8).

Compared with its one-hot variant, \sj{}-4B is stronger on JevBench
(JB-Hard 62.2 vs.\ 52.3; JB-All 80.5 vs.\ 75.8) and When2Call
(72.1 vs.\ 69.3), and holds the only best results among our models
(BANKING77 and CLINC150); the one-hot variant is higher on the Open-Jev
control tasks (64.5 vs.\ 57.0) and BoolQ (90.9 vs.\ 88.0).

\FloatBarrier

\section{Calibration and Decision Utility}
\label{app:proof}

The following result relates calibration to expected utility for a binary
decision. The comparison is with the best rule based on the model's own
scores, rather than a decision rule with additional information. Here
$\mathrm{ECE}(\rho)$ is an unbinned population quantity. The ECE values in
Tables~\ref{tab:main} and~\ref{tab:unseen} are finite-sample empirical
estimates and should not be substituted directly into this bound.

\begin{proposition}
\label{prop:regret}
Let $Y\in\{0,1\}$ be the outcome and $\rho\in[0,1]$ the model's predicted
probability of $Y=1$. Define the recalibration function
$g(z)=\Pr(Y=1\mid\rho=z)$.
For utility $U(a,y)\in[0,U_{\max}]$, let $a_p$ and $a_g$ be the
expected-utility-maximizing actions under $\mathrm{Bern}(\rho)$ and
$\mathrm{Bern}(g(\rho))$, respectively. Then
\[
  \mathbb{E}\big[U(a_g,Y)-U(a_p,Y)\big]
  \le 2\,U_{\max}\,\mathrm{ECE}(\rho),
\]
where $\mathrm{ECE}(\rho)=\mathbb{E}_{\rho}|g(\rho)-\rho|$ is the
population calibration error.
\end{proposition}

Thus, when the population calibration error is zero, the score-based rule
matches the best decision rule that can be constructed from the same score.
For nonzero error, the proposition bounds its expected utility gap.

\paragraph{Proof.}
For a score $z\in[0,1]$, write
$\bar U_z(a)=zU(a,1)+(1-z)U(a,0)$.
Conditioned on $\rho=z$, the true expected utility of action $a$ is
$\bar U_{g(z)}(a)$.
Since $U\in[0,U_{\max}]$, for every action $a$,
\[
  |\bar U_{g(z)}(a)-\bar U_z(a)|
  \le U_{\max}|g(z)-z|.
\]
At this score, $a_p$ maximizes $\bar U_z$, so
$\bar U_z(a_g)\le\bar U_z(a_p)$. It follows that
\begin{align*}
\bar U_{g(z)}(a_g)-\bar U_{g(z)}(a_p)
&\le \bar U_z(a_g)-\bar U_z(a_p) \\
&\quad + 2\,U_{\max}|g(z)-z|\\
&\le 2\,U_{\max}|g(z)-z|.
\end{align*}
Taking the expectation over the random score $\rho$ gives the claim. \qed

\FloatBarrier

\newpage

\section{Data and Model Licenses}
\label{app:licenses}

Table~\ref{tab:licenses} records official-release licenses checked on
1 October 2026. Datasets without a stated license are used only for
non-commercial research. Our query-assessment and query-document quality
sets are constructed for SearchDecision-Bench; source web pages remain
subject to their sites' terms.
\makeatletter
\setlength{\@dblfptop}{0pt}
\makeatother
\section{Experimental Settings}\label{app:Experimental Setting}
Table~\ref{tab:hparams-core} lists the full training and agent hyperparameters.

\begin{table*}[!t]
  \centering
  \scriptsize
  \setlength{\tabcolsep}{4pt}
  \begin{tabular}{@{}>{\raggedright\arraybackslash}p{0.19\textwidth}>{\raggedright\arraybackslash}p{0.28\textwidth}@{\hspace{10pt}}>{\raggedright\arraybackslash}p{0.19\textwidth}>{\raggedright\arraybackslash}p{0.28\textwidth}@{}}
    \toprule
    \multicolumn{2}{@{}l}{\textbf{\sj{} training (SLCD)}} & \multicolumn{2}{l@{}}{\textbf{Dual-system agent (BrowseComp-Plus)}} \\
    \cmidrule(r){1-2}\cmidrule(l){3-4}
    Backbone            & Qwen3.5-0.8B / 4B (instruct)         & System~2            & Qwen3.5-27B, thinking on \\
    Tuning              & LoRA, $r=16$, $\alpha=32$, dropout 0.05 & Sampling        & temperature 0.7, top-$p$ 0.8 \\
    Optimizer           & AdamW, lr $5{\times}10^{-5}$, wd 0.01 & Output budget       & 10{,}000 tokens per call \\
    Schedule            & cosine, 5\% warm-up                  & Iterations          & at most 100 per question \\
    Gradient clipping   & 1.0                                  & Context budget      & 128k tokens, then answer \\
    Precision           & bf16                                 & Retriever           & BM25, $n=50$ candidates \\
    Effective batch     & 132                                  & Snippet length      & 512 tokens \\
    Training steps      & 5,176 (1 epoch)                      & Relevance retention & $\tau_{\mathrm{keep}}=0.2$, 1--5 per search \\
    Max input length    & 2,048 tokens                         & Sufficiency         & more probable option \\
    Brier weight $\lambda_B$ & 0.5                             & Forced actions      & $\le$3 answers, $\le$2 continues \\
    Label weight $w_f$  & 0.5 for LLM-assigned labels, demonstrations, and rewrites; 1 otherwise & Sufficiency input & 6 documents, 200 tokens each \\
    Option permutation  & Choice and Noul fields               & Verification input  & 3 documents, 200 tokens each \\
    Temperature fitting & per output type, 5,000 validation decisions & Confidence gate & $\delta$= 0.8 \\
    Test cap            & 10,000 decisions per dataset         & Answer judge        & Qwen3-32B, temperature 0 \\
    Latency             & batch size 1, merged LoRA            & Questions / seed    & 100 / 42 \\
    \bottomrule
  \end{tabular}
  \vspace{-1mm}
  \caption{Core hyperparameters. \sj{} (one-hot) uses the same training
  settings with one-hot labels in place of soft targets; the AR (JSON)
  ablations in Table~\ref{tab:e2e} use the same agent settings.}
  \label{tab:hparams-core}
  \vspace{-2mm}
\end{table*}

\begin{table*}[!t]
  \centering
  \small
  \begin{tabular}{@{}l>{\raggedright\arraybackslash}p{12.8cm}@{}}
    \toprule
    License & Datasets and models \\
    \midrule
    Apache-2.0   & T2Ranking~\citep{xie2023t2ranking}, DuReader-Retrieval~\citep{qiu2022dureader}
                   and CBLUE~\citep{zhang2022cblue} (repository licenses),
                   CFEVER~\citep{lin2024cfever}, FRAMES~\citep{krishna2025frames},
                   XYZ-Aquila SFT, CValues~\citep{xu2023cvalues}
                   (repository license; data for research use),
                   typed-decisions;
                   Qwen3.5-0.8B/4B/27B~\citep{qwen2026qwen35} and
                   Qwen3-32B~\citep{yang2025qwen3} \\
    MIT          & RAGTruth~\citep{niu2024ragtruth}, TrendFact~\citep{zhang2026trendfact},
                   BrowseComp-Plus~\citep{chen2026browsecompplus}, JevBench,
                   Chinese Multi-Emotion Dialogue,
                   SemIf-OpenJev (code) \\
    CC BY 4.0    & MuSiQue~\citep{trivedi2022musique}, BANKING77~\citep{casanueva2020efficient},
                   MASSIVE~\citep{fitzgerald2023massive}, HelpSteer2~\citep{wang2024helpsteer2},
                   HelpSteer3~\citep{wang2025helpsteer3}, When2Call~\citep{ross2025when2call};
                   Search Arena~\citep{miroyan2026searcharena} prompts (model outputs
                   follow the providers' terms) \\
    CC BY 3.0    & CLINC150~\citep{larson2019evaluation} \\
    CC BY-SA 3.0 & QReCC~\citep{anantha2021qrecc}, VitaminC~\citep{schuster2021vitaminc}
                   (code: MIT), BoolQ~\citep{clark2019boolq} \\
    CC BY-SA 4.0 & SQuAD~2.0~\citep{rajpurkar2018know} \\
    CC0 1.0      & Open-Jev \\
    Mixed        & MultiNLI~\citep{williams2018broad} (OANC license, CC BY 3.0, CC BY-SA 3.0) \\
    Custom terms & PAWS~\citep{zhang2019paws} and PAWS-X~\citep{yang2019pawsx} (free use,
                   attribution to Google appreciated); MS MARCO~\citep{bajaj2016msmarco}
                   (non-commercial research only); Multilingual Amazon
                   Reviews~\citep{keung2020multilingual} (Amazon license terms);
                   THUCNews (free for research, commercial use by
                   agreement); AgentRewardBench~\citep{lu2025agentrewardbench} (research
                   use as fair use) \\
    Not stated   & Multi-CPR~\citep{long2022multicpr}, QBQTC,
                   LCQMC~\citep{liu2018lcqmc}, BQ Corpus~\citep{chen2018bq},
                   AFQMC~\citep{xu2020clue}, ATEC, the Rewrite corpus~\citep{su2019improving},
                   CANDY~\citep{guo2025candy}, CLUE tasks~\citep{xu2020clue} (CMNLI, OCNLI,
                   TNEWS, CSL, CLUEWSC, C3), ChnSentiCorp~\citep{tan2008empirical},
                   online\_shopping\_10cats and weibo\_senti\_100k,
                   LogiQA~\citep{liu2020logiqa}, COIG-P~\citep{coigp2025}, Douban and JD
                   reviews, kev-suites,
                   bev-decision (declared ``unknown'') \\
    \bottomrule
  \end{tabular}
  \vspace{-1mm}
  \caption{Licenses declared by the official releases of the datasets and
  models we use.}
  \label{tab:licenses}
  \vspace{-2mm}
\end{table*}

\FloatBarrier

\end{document}